\documentclass[12pt]{article}
\pdfoutput=1

\usepackage{hyperref}
\hypersetup{
  colorlinks=true,
  allcolors=darkpurple,
  pdfborder={0 0 0},
  linktocpage=false
}

\usepackage[normalem]{ulem}
\usepackage[utf8]{inputenc}
\usepackage[left=2.55cm, right=2.55cm, top=2.55cm, bottom=2.55cm]{geometry}
\usepackage{amsmath,amssymb}
\usepackage{slashed}
\usepackage{xcolor}
\usepackage{graphicx}
\usepackage{url}
\usepackage{cancel}
\usepackage{cite}
\usepackage{tabularx,booktabs}
\usepackage{multicol}
\usepackage{units}
\usepackage{xspace}
\usepackage[labelfont=bf]{caption}
\usepackage{mathrsfs}

\usepackage{breqn}
\usepackage{multirow}

\definecolor{darkred}{rgb}{0.6,0,0}
\definecolor{darkpurple}{rgb}{0.5,0,0.5}

\def\gsim{\raise0.3ex\hbox{$\;>$\kern-0.75em\raise-1.1ex\hbox{$\sim\;$}}}
\def\lsim{\raise0.3ex\hbox{$\;<$\kern-0.75em\raise-1.1ex\hbox{$\sim\;$}}}

\renewcommand{\arraystretch}{1.6}

\begin{document}

\vspace*{-2cm}

\begin{center}
\vspace*{15mm}

\vspace{1cm}
{\Large \bf 
 Quasi-Dirac neutrinos in the linear seesaw model } \\
\vspace{1cm}

{\bf Carolina Arbel\'aez$^{\text{a}}$, Claudio Dib$^{\text{a}}$,
 Kevin Mons\'alvez-Pozo$^{\text{b}}$, Iv\'an Schmidt$^{\text{a}}$  }

\vspace*{.5cm} $^{\text{a}}$ Universidad T\'ecnica Federico Santa Mar\'ia and Centro Cient\'ifico Tecnol\'ogico
de Valpara\'iso CCTVal, \protect\\ Avenida Espa\~na 1680,
Valpara\'iso, Chile

\vspace*{.5cm} $^{\text{b}}$Instituto de F\'isica Corpuscular, CSIC -- Universitat de Val\`encia,\\ Apt.\ Correus 22085, E-46071 Val\`encia, Spain

 \vspace*{.3cm}
 \href{mailto:carolina.arbelaez@usm.cl}{carolina.arbelaez@usm.cl},
 \href{mailto:claudio.dib@usm.cl}{claudio.dib@usm.cl},
 \href{mailto:kevin.monsalvez@ific.uv.es}{kevin.monsalvez@ific.uv.es},
 \href{mailto:ivan.schmidt@usm.cl}{ivan.schmidt@usm.cl}
\end{center}

\vspace*{10mm}
\begin{abstract}\noindent\normalsize
We implement a minimal linear seesaw model (LSM) for addressing the Quasi-Dirac (QD) behaviour of heavy neutrinos, focusing on the mass regime of $M_{N} \lesssim M_{W}$. 
Here we show that for relatively low neutrino masses, covering the few GeV range, the same-sign to opposite-sign dilepton ratio, $R_{\ell  \ell}$, can be anywhere between 0 and 1, thus signaling a Quasi-Dirac regime. Particular values of $R_{\ell \ell}$ are controlled by the width of the QD neutrino and its mass splitting, the latter being equal to the light-neutrino mass $m_{\nu}$ in the LSM scenario. The current upper bound on $m_{\nu_{1}}$ together with the projected sensitivities of current and future $|U_{N \ell}|^{2}$ experimental measurements, set stringent constraints on our low-scale QD mass regime. Some experimental prospects of testing the model by LHC displaced vertex searches are also discussed.

\end{abstract}

\section{Introduction\label{sect:intro}}

Neutrino oscillation experiments have found overwhelming evidence for the existence of non vanishing --albeit small-- neutrino masses \cite{Fukuda:1998mi,Ahmad:2002jz,deSalas:2020pgw}, thus showing the existence of Physics Beyond the Standard Model. Despite the accuracy of the experiments, fundamental theoretical questions still remain, for instance the one regarding the Dirac vs. Majorana nature of the neutrinos. The most widely accepted mechanism to generate small neutrino masses is the well-known seesaw mechanism \cite{MINKOWSKI1977421,PhysRevLett.44.912,PhysRevD.22.2227,PhysRevD.22.2860}, which involves extra heavy neutral fermions,  denoted here as $N_{i}$ ($i=1,2,\ldots, n$, depending on the model). 
In most of these scenarios, the heavy neutrinos are Majorana fermions. 
However, there are seesaw scenarios where pairs of these Majorana neutrinos approach smoothly their mass degeneracy limit, 
$\Delta M_N \to 0$.  In the exact degeneracy limit ($\Delta M_N=0$) these neutrinos become Dirac fermions and the lepton number violating (LNV) processes they induce cancel exactly. Instead, in  the approximate degeneracy case, 
when $\Delta M_N$ is small but still finite (comparable to $\Gamma_N$),  the LNV processes they induce cancel only partially. These almost degenerate Majorana neutrinos are usually called \emph{Quasi-Dirac} neutrinos. These Quasi-Dirac neutrinos could appear at scales not far below the electroweak scale in some extensions, such as in the inverse seesaw \cite{PhysRevD.34.1642, Khalil:2010iu, Das:2014jxa, Das:2015toa, Anamiati:2016uxp,Das:2017hmg} and in the linear seesaw \cite{Akhmedov:1995ip, Malinsky:2005bi, Dib:2014fua}. 
In this article, we will work within the framework of the minimal linear seesaw model, which naturally yields pairs of Quasi-Dirac right-handed neutrinos $N$ and $N^{\prime}$ in a regime of masses below $M_W$.

While lepton number conserving (LNC) processes are mediated by either Dirac or Majorana neutrinos, lepton number violating processes are induced by Majorana neutrinos only. This feature is used in experimental searches to discriminate the nature of the hypothetical heavy neutrinos.
At hadron colliders, a clear signal of Majorana neutrinos would come from the same-sign (\emph{SS}) dileptons produced via LNV processes $pp \rightarrow W \rightarrow \ell^{\pm} N\rightarrow \ell^{\pm} \ell^{\pm}jj$. These modes involve two jets and no missing transverse energy. Similar events with opposite-sign (\emph {OS}) lepton pairs could also be produced through $pp \rightarrow W \rightarrow \ell^{\pm} N \rightarrow \ell^{\mp} \ell^{\pm}jj$~\cite{Gluza:2015goa}; however, these are LNC processes, and so can be mediated by both Dirac and Majorana neutrinos. Searches of these $\ell \ell jj$ events, for heavy neutrinos with mass above $M_W$, have been done at ATLAS and CMS \cite{Aad:2015xaa,Khachatryan:2015gha,CMS:2018szz}. 

For masses below $M_{W}$, the jets may not be energetic enough to be separated from the background, so searches of trileptons events $\ell^{\pm}\ell^{\pm}\ell^{\prime\mp}\nu$ could provide a more favorable signal \cite{Izaguirre:2015pga,delAguila:2008hw,delAguila:2009bb,Chen:2011hc,Dube:2017jgo,CMS:2018iaf}.
However, here the lepton charges cannot tell us about the Dirac or Majorana nature of the intermediate neutrinos.
Several ways to distinguish between Dirac or Majorana neutrinos using  these  pure  leptonic  modes  at the  LHC  have  been proposed \cite{Dib:2016wge, Dib:2017iva, Dib:2017vux, Arbelaez:2017zqq}. An alternative for distinguishing Dirac from Majorana neutrinos at the LHC, with masses in the range $5$ GeV $< M_N<20$ GeV, have also been proposed \cite{Dib:2019ztn, Dib:2018iyr}, now using the exclusive semileptonic processes $W \rightarrow  \ell (N\rightarrow \ell \pi, \ell 2\pi, \ell 3\pi)$, which again have no missing energy. Other ways to resolve the nature of neutrinos are given in Refs.~\cite{Balantekin:2018ukw,deGouvea:2021ual,Blondel:2021mss}.

Going back to the $\ell\ell j j $ modes, 
the ratio of SS to OS events, which we call $R_{\ell\ell}$, will indicate the Majorana/Dirac nature of the intermediate neutrinos that induce these events~\cite{Gluza:2016qqv}. As mentioned above, Majorana neutrinos induce SS and OS in equal amount, thus $R_{\ell\ell}=1$ in that case. In contrast, Dirac neutrinos only induce OS events, thus $R_{\ell\ell}=0$.

A measurement of $R_{\ell \ell}$ different from zero or unity provides then valuable information about the Dirac/Majorana character of the heavy neutrinos, and thus about the mechanism underlying the neutrino mass generation. Indeed, in an inverse seesaw scenario with Quasi-Dirac neutrinos \cite{Anamiati:2016uxp, BhupalDev:2015kgw}, it was pointed out that dilepton states will exhibit $R_{\ell \ell} \neq 0,1$, where the precise value $0<R_{\ell \ell}<1$ is controlled by the heavy neutrino mass splitting (which in the inverse seesaw scenarios is proportional to the small  parameter $\mu_{R}$ that softly breaks lepton number) and the neutrino decay width. In Ref.  \cite{Das:2017hmg}, it is also pointed out that values $0<R_{\ell \ell}<1$ are possible as long as there exists a \emph{CP} violating phase. 
 
Similar to the inverse seesaw scenario, in the linear seesaw another softly-breaking lepton-number mass parameter, denoted here as $M_{\epsilon}$, also allows to naturally fit small masses for the light neutrinos, without requiring GUT scale masses for the heavy states. However, in the linear seesaw the mass splitting of the heavy states equals $m_\nu$ (the light neutrino mass), unlike the inverse seesaw where the splitting is given by the parameter $\mu_R$. This feature will imply a different mass range for the Quasi-Dirac regime: in the linear seesaw scenario, which we study here, $R_{\ell \ell}$ takes values different from 0 or 1 at lower $M_N$, including a few GeV. 
 Therefore, sterile neutrino searches at current and near-future experiments such as SHiP, ANUBIS, MATUSHLA and DUNE \cite{Deppisch:2015qwa,deVries:2020qns,Hirsch:2020klk}, set interesting constraints on the parameter space of the linear seesaw scenario. 
 It is worth to highlight that a measurement of $R_{\ell \ell}$ requires both leptons to be detected and their charge to be identified. Such low mass experiments can not detect these signals. However, ATLAS and CMS can probe this scenario by displaced vertex searches, covering neutral lepton masses roughly within 10-30 GeV \cite{Cottin:2018nms, Aaboud:2017iio, Aad:2015rba, Zhu:2020lww}. Some experimental prospects for testing Quasi-Dirac neutrinos in the linear seesaw model, coming from displaced vertex searches, are also briefly discussed in this work.

It is important to note that, since values of $M_N < 2.5$ GeV enter in our analysis, an appropriate treatment of non-perturbative QCD has to be considered for
the heavy neutrino decay calculations. In this work we use the formulation of the \emph{resonance chiral theory} \cite{Ecker:1988te,Ecker:1989yg,Cirigliano:2006hb}. 
Also, a special and convenient parametrization of seesaw scenarios, described in \cite{Cordero-Carrion:2019qtu}, was used here in order to fit the light-neutrino data and also to properly relate input parameters with Quasi-Dirac and experimental constraints.

The text is organized as follows. In Section \ref{sect:model}, we recall the main features of the linear seesaw and discuss some relations on the neutrino masses appearing in linear seesaw scenarios. In Section \ref{sect:sum}, we show the $R_{\ell \ell}$ definition in the limit where the decay widths of both Quasi-Dirac neutrinos are of the same order. A detailed description of the total decay width of the heavy neutrino, including the hadronic decays calculated in the non-perturbative QCD regime $M_N<2.5$ GeV, is done in Section \ref{sect:DecayRates}. In Section \ref{sect:results}, we describe the Yukawa parametrization based on two simple and convenient ansatzes for the input matrices. Here we also discuss the relevant results concerning the constraints on the parameter space coming from the Quasi-Dirac condition $0<R_{\ell \ell}<1$. Restrictions on the $M_{N}$ values, coming from the low-mass experimental bounds and also from the sensitivity of the displaced vertex searching at the LHC, are also depicted in this Section. We close with a short summary and conclusions in Section \ref{sect:conc}.

\section{Model setup\label{sect:model}}
\noindent In this section we discuss the main aspects of the linear seesaw mechanism (LSM). Up to the SM, the minimal version of the LSM contains two different types of neutral $SU(2)$ singlet fermions $(N,S)$ per generation. In addition to the kinetic sector, the Lagrangian contains the following  terms:   

\begin{equation}
\label{eq::YukLag}
    \mathcal L_{Y} = Y_{D} \overline{L} H^{c} N + Y_{\epsilon}  \overline{L}H^{c} S + M_{R} \overline{N^{c}} S + \text{h.c.}
\end{equation}

We have omitted flavor indices to simplify the notation. In three generations, $Y_{D}$ and $Y_{\epsilon}$ are $3 \times 3$ Yukawa matrices, with $Y_{D} \neq Y_{\epsilon}$, and $M_{R}$ is a $3 \times 3$ complex matrix. In SM-seesaw extensions one can always perform a change of basis, in this case on $N$ and $S$, such that $M_{R}$ becomes diagonal. Here we take $M_{R}=\text{diag}(M_{R_{1}},M_{R_{2}},M_{R_{3}})$.

Considering the basis in Eq.~\eqref{eq::YukLag}, namely $(\nu_{L}^c,N,S)$, the texture of the neutrino mass ($9 \times 9$) matrix, given in a $3 \times 3$ block notation, reads:

\begin{equation}
    M_{\nu}=\begin{pmatrix} 
    0 & m_{D} & M_{\epsilon}\\
    m_{D}^{T} & 0 & M_{R}\\
   M_{\epsilon}^{T} & M_{R}^{T} & 0
    \end{pmatrix},
\label{eq:numatrix}
\end{equation}

\noindent where $m_{D}= v_{SM} Y_{D}/ \sqrt{2}$ and $M_{\epsilon} = v_{SM} Y_{\epsilon}/ \sqrt{2}$. Bringing this $3 \times 3$ matrix into a block-diagonal form ---through a diagonalization-like procedure considering $ M_{\epsilon} \ll m_{D} < M_{R}$--- the non-diagonal mass matrix of the light neutrinos is given by
\begin{equation}
    m_{\nu}= m_{D} M_{R}^{-1} M_{\epsilon} ^{T} + M_{\epsilon} M_{R}^{T^{-1}} m_{D}^{T}  = \frac{v^2}{2} (Y_{D} M_{R}^{-1} Y_{\epsilon}^{T} + Y_{\epsilon} M_{R}^{T^{-1}}Y_{D}^{T}) .
    \label{eq:mnu}
\end{equation}
The analogous expressions for the two $3 \times 3$ mass matrices of the heavy neutrinos are

\begin{equation}
\label{eq:MN12}
M_{N_{a},\, N_{b}}\simeq \frac{M_{R}}{2}+\frac{m_{D}^{2}M_{R}^{-1}}{4}\mp \frac{m_{D}M_{R}^{-1}M_{\epsilon}^{T}}{2} + \text{h.c.}
\end{equation}

From Eq.~\eqref{eq:mnu}, the effective $\nu$ mass is roughly directly proportional to $m_{D}$, also proportional to $M_{\epsilon}$, and inversely proportional to $M_{R}$. The smallness of the lepton-number violating term $M_{\epsilon}$, together with the linearity on $m_{D}$, gives this model the name \emph{low-scale linear seesaw}, and generates 
small $m_{\nu}$ masses  for the light neutrinos, without requiring $M_{R}$ to be extremely large, beyond the reach of any foreseeable experiment. This is why we can deal with heavy neutrinos states in the region of masses of the few GeV's. Note that although $N_{a}$ and $N_{b}$ are not in general the heavy-neutrino mass eigenstates, which are really obtained from the diagonalization of the full $9 \times 9$ matrix in Eq.~\eqref{eq:mnu}, due to the large mass gap between the light and heavy neutrinos one can identify Eq.~\eqref{eq:MN12} as the actual heavy-neutrino mass matrix in the region of interest of the parameter space of the model.

The Lagrangian, extended with the heavy neutral leptons (HNL) of Eq.~\eqref{eq::YukLag}, turns the SM-flavour neutrinos 
 (\emph{i.e.} those that couple in the charged lepton currents)
 into a mixture of the light and heavy mass eigenstates. For a given flavour $\ell\,$,
\begin{equation}
\label{eq:neutrinomixing}
\nu_{\ell}=\sum_{k=1}^{3}U_{\ell \nu_{k}}\nu_{k}+\sum_{k=1}^{3}U_{\ell N_{k}}N_{k}+\sum_{k=1}^{3}U_{\ell N^{\prime}_{k}}N^{\prime}_{k}\, ,
\end{equation}
where $\nu_{k}$ are the three light neutrino mass eigenstates, $U_{\ell \nu_{k}}$ is the Pontecorvo-Maki-Nakagawa-Sakata (PMNS) matrix and $N$ and $N^{\prime}$ stand now for the two heavy-neutrino mass eigenstates for each generation (with $k=1,2,3$ denoting the generation). Note that the PMNS matrix is not unitary anymore: unitarity is only preserved for the more general $9\times 9$ mixing matrix $U$. However, current data excludes large departures from unitarity of the PMNS matrix, which entails small mixings among active and heavy neutrinos, $U_{\ell N}$.

An important feature of the LSM is that the mass splitting between the two heavy neutrinos within each generation is very small: 
\begin{equation}
    \Delta M_{i}\sim m_{\nu_{i}}\, ,
    \label{eq:splitt}
\end{equation}
as it can be easily seen from~\eqref{eq:mnu} and~\eqref{eq:MN12}. This is crucial in order to study the Quasi-Dirac behaviour of pairs of heavy neutrinos in the linear seesaw model, as it is thoroughly described in the next section.

\section{Quasi-Dirac Neutrinos\label{sect:sum}}
The usual Dirac-Majorana dichotomy regarding the nature of neutrinos is somehow misleading, since the Dirac case can be considered as a limiting case of a more general Majorana scenario with twice the neutrino content. Indeed, a single Dirac neutrino corresponds to a pair of Majorana neutrinos exactly degenerate in mass, so from a scenario of $2n$ Majorana neutrinos one reaches a scenario of $n$ Dirac neutrinos if the Majorana masses become degenerate in pairs. At that point, all sources of lepton number violation in the model vanish.  Now, when this Dirac limit is reached in a continuous way --- by gradually switching off the LNV mass terms --- one crosses an interesting, albeit narrow regime, usually called \emph{Quasi-Dirac}.

An observable that is commonly used at the LHC to look for Majorana neutrinos is the same-sign to opposite-sign dilepton ratio in $\ell\ell jj$ events with no missing $p_T$, which we call  $R_{\ell\ell}$. 
The production of these events with a pair of leptons of the same sign, are expected to occur only through lepton number violating processes mediated by Majorana neutrinos, namely $\overline{q}q \rightarrow W^{\pm} \rightarrow \ell_{\alpha} ^{\pm} N^{(\prime)} \rightarrow \ell_{\alpha} ^{\pm}l_{\beta} ^{\pm}W^{\prime \mp}$, where $\overline{q}$ and $q$ denote the partons inside the colliding protons, while $N^{(\prime)}$ denotes the two heavy-neutrino mass eigenstates \footnote{The $\ell$-flavor $N_{\ell}$ heavy state and its conjugated $(N_{\overline{\ell}})$ produced in the decays $W^{+} \rightarrow \overline{\ell}N_{\ell}$, $(W^{-} \rightarrow \ell N_{\overline{\ell}})$ can be written in terms of the QD mass eigenstates as: $N_{\ell}= (N - i N^{\prime})/\sqrt{2}$,  $N_{\overline{\ell}}= (N + i N^{\prime})/\sqrt{2}$.}. 
 On the other hand, opposite-sign pairs of leptons are produced via lepton number conserving processes mediated by both Dirac and Majorana neutrinos, as $\overline{q}q \rightarrow W^{\pm} \rightarrow \ell_{\alpha} ^{\pm} N^{(\prime)} \rightarrow \ell_{\alpha} ^{\pm}\ell_{\beta} ^{\mp}W^{\prime\pm} $. 
The virtual $W'$ gauge boson then turns either into $q\bar q'$ 
producing the final state $\ell\ell j j$ or into $\ell\nu$ producing a trilepton plus missing energy.
For $M_N \gtrsim 5$ GeV  only around the 40\% of the total number of events are into $\ell\ell j j$,
which are those used for measuring the SS to OS dilepton ratio $R_{\ell \ell}$.

It is important to highlight the fact that prompt searches of such charged leptonic signals are background dominated, while the displaced vertex (DV) events are background free. Therefore, a more favorable measurement of $R_{\ell \ell}$ through $\ell\ell jj$ signals involves the detection of ``displaced dileptons" plus jets. The region of parameters sensitive to such DV searches is described in the next sections.

 The ratio of SS over OS events between times $t_{a}$ and $t_{b}$, after heavy-neutrino production, is given by the ratio of the time-integrated amplitudes squared as \cite{Anamiati:2016uxp,Das:2017hmg,Antusch:2017ebe}:

\begin{equation}
R(t_{a},t_{b})=\frac{\int_{t_{a}}^{t_{b}} |g_{-}(t)|^{2} dt}{\int_{t_{a}}^{t_{b}} |g_{+}(t)| ^{2} dt},
\end{equation}
where $g_{-}(t) \sim -i e^{-iMt} e^{-\frac{\Gamma}{2}t}\sin(\frac{\Delta M}{2}t)$ and $g_{+}(t) \sim -ie ^{-iMt} e^{-\frac{\Gamma}{2}t}\cos(\frac{\Delta M}{2}t)$ are the oscillating amplitudes, with $M=\frac12 (M_{N}+M_{N^{\prime}})$ and  $\Delta M= M_{N}-M_{N^{\prime}}$, and $\Gamma$ is the total decay width of both heavy neutrino states ($\Gamma = \Gamma_{N} \sim \Gamma_{N^{\prime}}$).

Oscillations between two quantum mechanical states can occur whenever these states can be distinguished by some quantum number (which is not conserved, hence the oscillation); however, their mass eigenstates are admixtures of them. In the present case the mass eigenstates $N$ and $N'$ are admixtures of $N$ and $S$, which can be assigned opposite lepton number. Then $|g_{-}(t)|^{2}$ describes the probability of a lepton number conversion in the time interval $t$, while $|g_{+}(t)|^{2}$ denotes the probability 
that lepton number is not converted in that time interval. Sometimes this oscillation is described as between neutrino and antineutrino, however this denomination may lead to confusion, since a Majorana neutrino is its own antiparticle; the essential point is that there are two (or more) neutrinos with different mass and that there is a non-conserved quantum number that characterizes the states that oscillate, which are different than the mass states.

\begin{figure}[t]
\centering
\includegraphics[width=0.49\textwidth]{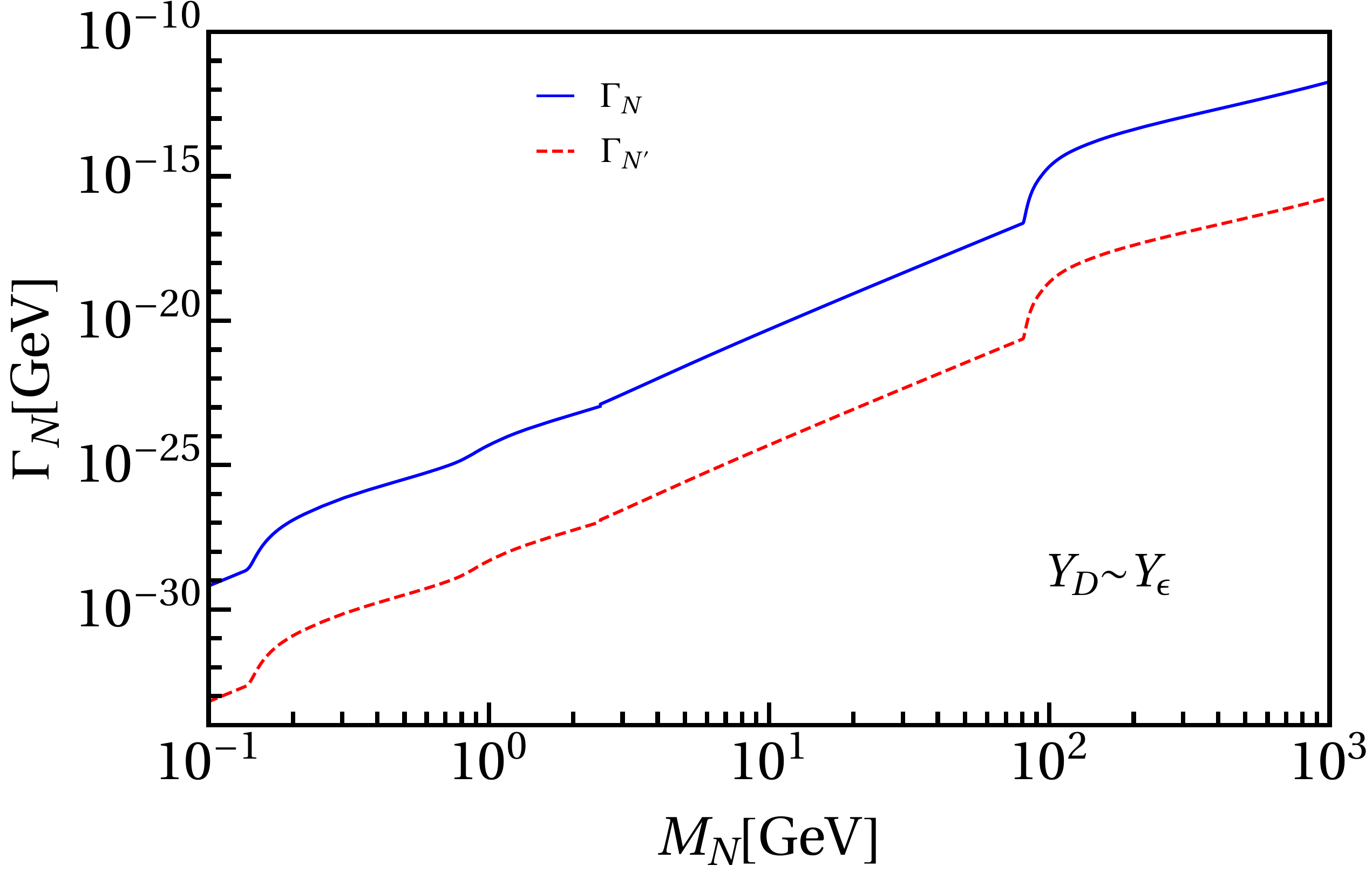}
\includegraphics[width=0.49\textwidth]{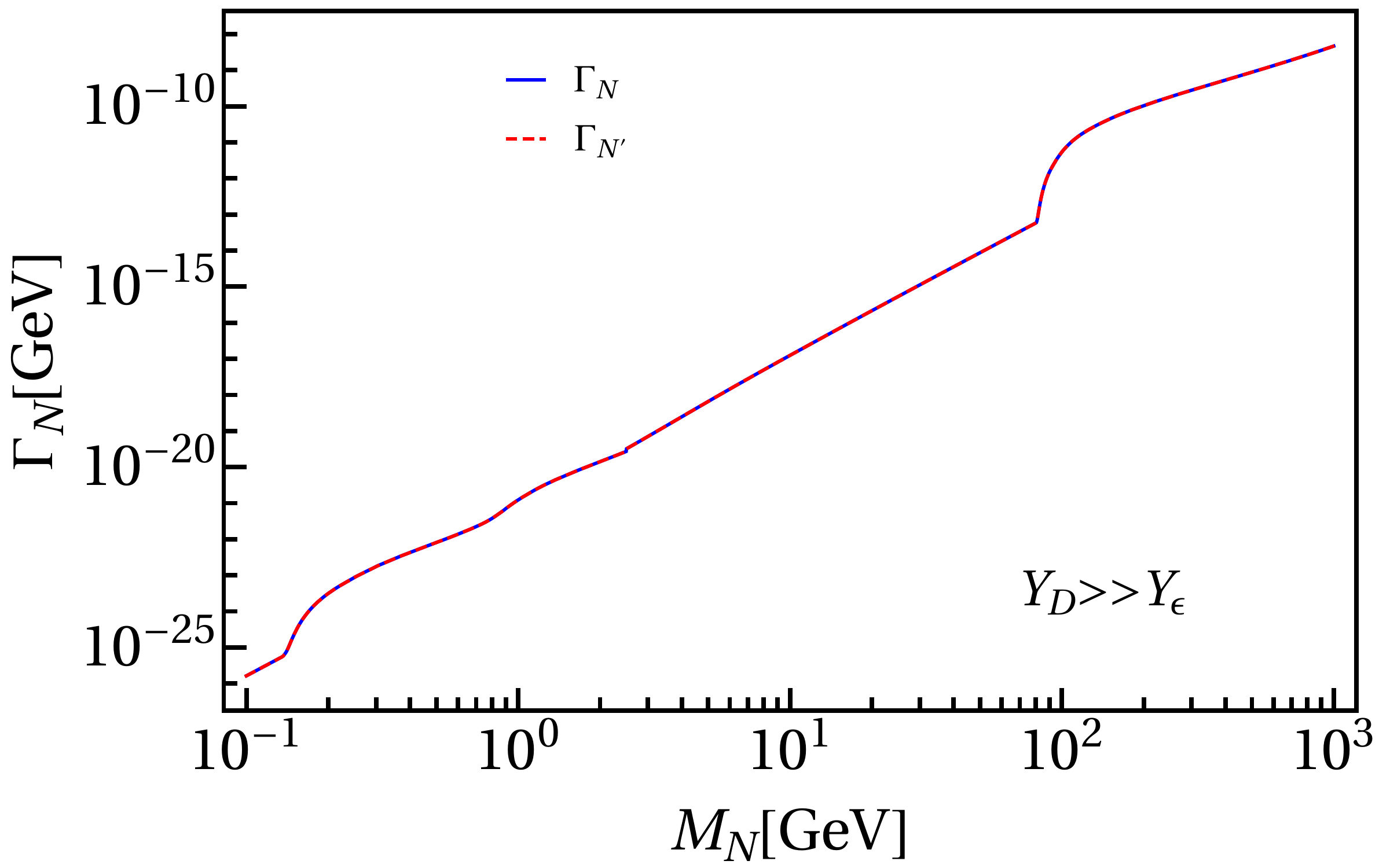}
\caption{$\Gamma_{N^{(\prime)}}$ vs $M_{N^{(\prime)}}$. Left and right panels correspond to the cases where $Y_{\epsilon} \sim  Y_{D}$  and  $Y_{\epsilon} \ll Y_{D}$ respectively. Both are shown within \emph{Scenario b} for $g=1$ (left) and $g=10^{2}$ (right), see Section~\ref{sect:results} for more details. }
    \label{fig:figGamma}
\end{figure}

The approximations used for the probabilities are only valid in the limit where the decay widths of both Quasi-Dirac neutrinos are approximately equal, i.e:  $\Delta \Gamma = \Gamma_{N}-\Gamma_{N^{\prime}}\to 0$ \cite{Antusch:2017ebe}. In terms of the Yukawa couplings of our model, $\Gamma_{N}\sim \Gamma_{N^{\prime}}$ as long as $Y_{\epsilon} \ll Y_{D}$. In an attempt to scrutinize this aspect, we show in Fig.~\ref{fig:figGamma} the $\Gamma_{N^{(\prime)}}$ decay widths versus $M_{N^{(\prime)}}$ for the case where $Y_{\epsilon}=Y_{D}$ (left panel) and $Y_{\epsilon}= 10^{-4} Y_{D}$ (right panel). It is clearly appreciated that the larger the hierarchy among the Yukawas, the more accurate is the relation $\Gamma_{N} \sim \Gamma_{N^{\prime}}$ and then, the better the approximations of the probabilities $g_{\pm}(t)$ used here. In the next sections, we highlight the impact of the hierarchy of the Yukawa couplings in our numerical results.

For our analysis, we consider the ratio $R_{\ell \ell}=R(0,\infty)$, which can be expressed as:

\begin{equation}
R_{\ell \ell}=\frac{\Delta M^{2}}{2 \Gamma^{2}+\Delta M^{2}}\, .
\label{eq:Rll}   
\end{equation}
Note that $R_{\ell \ell} \rightarrow 1$ if $\Gamma \ll \Delta M$,  
while $R_{\ell \ell} \rightarrow 0$ if $\Delta M \ll \Gamma$. 
These two scenarios correspond to the limiting Majorana and Dirac cases, respectively. In models with Quasi-Dirac neutrinos, the ratio $R_{\ell \ell}$ can take any value between $0$ and $1$; as we approach $R_{\ell \ell} \to 0$, the LNV effects are more and more suppressed. From Eq.~\eqref{eq:Rll}, the QD condition $R_{\ell \ell} \neq 0,1$ is ensured as long as $\Gamma (N) \sim \Delta M$. Using the estimate for the mass splitting from Eq.~\eqref{eq:splitt}, it follows that

\begin{equation}
\Gamma (N) \sim m_{\nu}\, . 
\end{equation}

Therefore, the window of $R_{\ell \ell}$ values compatible with QD neutrinos is determined, in the linear seesaw model, by the light-neutrino masses and the heavy-neutrino decay width. In the next sections, we present first the decay modes of the heavy neutrinos, focusing on the hadronization of the final-state quark currents within the non-perturbative QCD regime below $\sim 2.5$ GeV, and second, we explore in detail the parameter space provided by the above conditions and its phenomenological implications.

\section{Heavy-neutrino decay rates}
\label{sect:DecayRates}

In order to characterize the Quasi-Dirac regime in the linear seesaw model, one needs to compute the total decay width of the heavy neutrino, $\Gamma$, that appears in Eq.~\eqref{eq:Rll}. The mixture of the SM-flavour neutrinos in Eq.~\eqref{eq:neutrinomixing} provides the interactions of the --otherwise sterile-- heavy neutrinos with the SM fields, i.e.
\begin{equation}
\label{eq::IntLag}
\mathcal{L}_{\text{int}}=\frac{g}{2 \sqrt{2}}W_{\mu}^{\dagger}\bar{N^{c}}\sum_{\alpha}U_{\alpha}^{\ast}\gamma^{\mu}(1-\gamma_{5})\ell_{\alpha}^{-}+\frac{g}{4 \cos \theta_{W}}Z_{\mu}\bar{N^{c}}\sum_{\alpha}U_{\alpha}^{\ast}\gamma^{\mu}(1-\gamma_{5})\nu_{\alpha}+\text{h.c.}\, ,
\end{equation}
with $\theta_{W}$ the weak (or Weinberg) angle. 
Hence, all the decays of the heavy neutrinos proceed via charged and neutral-current interactions.
We will follow the usual considerations and assume that the three families of the heavy neutrinos are not degenerate in mass, so that each leaves a different footprint in the experiment. We will assume only one pair of heavy neutrinos to be relatively light, while the other two pairs to be much heavier, namely  $M_{N_{1}^{(\prime)}} \ll M_{N_{2}^{(\prime)}} \lesssim M_{N_{3}^{(\prime)}}$. Thus, we will focus on the lightest heavy neutrino pair, which for some mass range can be within the expected sensitivities of several experiments, compute their decay width and set constraints on the parameter space.

We are interested in the regime where $\Gamma_{N_{1}} \sim \Gamma_{N_{1}^{\prime}} $, 
so we just need to calculate one of these decay widths for the $R_{\ell \ell}$ ratio of Eq.~\eqref{eq:Rll}. The width necessarily depends on the HNL mass, since larger masses open more decay modes. We will separate these modes into four categories based on the nature of the final states~\cite{Atre:2009rg,Helo:2010cw,Bondarenko:2018ptm,Cvetic:2015naa}:

\begin{center}
\begin{tabular}{ c c c c }
Leptonic decays & Semileptonic decays & Decays into gauge bosons & Decay into the Higgs boson \\ 
\llap{\textbullet} $N\rightarrow \ell^{-}\ell^{\prime \,  +} \nu_{\ell^{\prime}}$ & \llap{\textbullet} $N\rightarrow \ell^{-}u\bar{d}$ & \llap{\textbullet} $N\rightarrow \ell^{-} W_{L}^{+}$  & \llap{\textbullet}$N\rightarrow \nu_{\ell} H $  \\
\llap{\textbullet} $N\rightarrow \nu_{\ell} \ell^{\prime \, -}\ell^{\prime \,  +} $ & \llap{\textbullet} $N\rightarrow \nu_{\ell} q \bar{q} $ & \llap{\textbullet} $N\rightarrow \ell^{-} W_{T}^{+}$ &  \\
\llap{\textbullet} $N\rightarrow \nu_{\ell} \nu \nu$ & & \llap{\textbullet}$N\rightarrow \nu_{\ell} Z_{L} $ & \\
& &\llap{\textbullet}$N\rightarrow \nu_{\ell} Z_{T} $  & 
\end{tabular}
\end{center}

Here, $L$ and $T$ denote the longitudinal and transverse polarizations of the gauge bosons. The $u$ and $d$ stand for 
the up and down-type quarks respectively, while $q$ includes all the quark types
that are kinematically accessible in the decays\footnote{The contributions from the different channels into top quarks are not considered here, since they are negligible as compared to the decay into bosons.}. The first three columns come from the interaction Lagrangian of Eq.~\eqref{eq::IntLag}, either from charged-current or neutral-current interactions, whereas the Higgs decay is directly driven by the Yukawa Lagrangian. Note that for Majorana neutrinos, one should also consider the charge conjugated mode of the charged-current interacting processes, thus summing twice to the total widths.

The treatment of the decay modes of the heavy neutrino into quarks entails an inherent difficulty due to the non-perturbativity of QCD for $M_{N_{1}}\lsim 2.5$ GeV. While perturbation theory works properly for larger masses of the heavy neutrino and, consequently, the usual computation of partial decays into quarks is appropriate to consider in the inclusive decays, in the non-perturbative regime below $\sim 2.5$ GeV we are forced to consider the hadronization of the final quark currents into exclusive hadrons, which are the relevant degrees of freedom at those energies. For this purpose, a model-independent scheme is provided by chiral perturbation theory ($\chi$PT)~\cite{Gasser:1983yg,Gasser:1984gg}. This allows us to handle the hadronization procedure of quark currents into mesons. However, this framework alone is not enough to describe the whole non-perturbative energy regime, since it gives reliable results only for $E \ll 1$ GeV. The intermediate region is populated by hadron resonances, and therefore, a complementary scenario that encompasses these resonances as well as all the features of $\chi$PT, is given by the resonance chiral theory (R$\chi$T)~\cite{Ecker:1988te,Ecker:1989yg,Cirigliano:2006hb}. We refer the reader to~\cite{Husek:2020fru} for a detailed explanation of the hadronization procedure and its limitations within R$\chi$T. 

Accordingly, we will consider the hadronization of the quark bilinears into one pseudoscalar ($P$), two pseudoscalars ($PP$) or a vectorial resonance ($V$) ---considering all possible intermediate states provided by R$\chi$T--- with only light-quark content, i.e. $u, d$ and $s$. The list of final states, as resulting either from a neutral or charged current interactions, are shown in Table~\ref{tab::FinalHadronicStates}. A last remark before presenting the results is in order: due to the fast decay of hadron resonances, these are not proper asymptotic states, as it is required in quantum field theory, and consequently, the decay width of the heavy neutrino into a vectorial resonance as an external state is not well defined. In order to compute these decays we use the method employed in Refs.~\cite{Arganda:2008jj,Lami:2016vrs,Husek:2020fru}, based on the decay of these resonances into two pseudoscalars (see the parentheses in the third column of Table~\ref{tab::FinalHadronicStates}), i.e.
\begin{equation}
\label{eq:brconst2}
\mathcal{B}(N \rightarrow \ell (\nu) V) =    \sum_{P_1 P_2} \, \mathcal{B}(N \rightarrow \ell (\nu) P_1 P_2)\Big|_V \,.
\end{equation}
This treatment applies to the resonances that decay fast and mainly into two pseudoscalars. The $\omega(782)$ is not of this kind and is thus treated as a well-behaved external state.
\begin{table}[h!]
\begin{center}
\renewcommand{\arraystretch}{1.5}
\begin{tabular}{|c||c|c||c|c||c|}
\cline{2-6}
\multicolumn{1}{c|}{} & \multicolumn{2}{c||}{$PP$} & \multicolumn{2}{c||}{$P$} & \multicolumn{1}{c|}{$V$} \\
\cline{2-6}
\cline{1-6}
\multirow{6}{8.9em}{Neutral-current interactions \\ ($N\rightarrow \nu$ hadron(s))} & \multicolumn{2}{c||}{\multirow{2}{*}{$ \pi^{+}\pi^{-}$}}  & \multirow{2}{*}{$\pi^{0}$} & \multirow{3}{*}{$ \eta$} & \multirow{2}{*}{$\rho^{0}\,(\pi^{+}\pi^{-})$} \\
	&	\multicolumn{1}{c}{}  &	&	&	&	\\
	\cline{2-4}     \cline{6-6}
	& \multicolumn{2}{c||}{\multirow{2}{*}{$ K^{0}\overline{K}^{0}$}} &  \multirow{2}{*}{$ K^{0}$} &   & \multirow{2}{*}{$ \phi\, (K^{0}\overline{K}^{0}+K^{+}K^{-})$} \\
				\cline{5-5}
	&\multicolumn{1}{c}{} 	&	&	& \multirow{3}{*}{$ \eta^{\prime}$}	&	\\
	\cline{2-4}		\cline{6-6}
 &  \multicolumn{2}{c||}{\multirow{2}{*}{$K^{+}K^{-}$}}  & \multirow{2}{*}{$ \overline{K}^{0}$} & & \multirow{2}{*}{$ \omega$}   \\
	&	\multicolumn{1}{c}{} &	&	&	&	\\
\hline
\hline
\multirow{12}{8.9em}{Charged-current interactions \\ ($N\rightarrow \ell^{\pm}$ hadron(s))}  &  \multirow{2}{*}{$\pi^{-}\pi^{0}$}  & \multirow{2}{*}{$\overline{K}^{0} \pi^{-}$} &  \multicolumn{2}{c||}{\multirow{3}{*}{$\pi^{-}$}} & \multirow{3}{*}{$\rho^{-}\,(\pi^{-}\pi^{0})$}   \\
	&	&	&	\multicolumn{1}{c}{} &	&	\\
\cline{2-3}
 & \multirow{2}{*}{$ K^{0} K^{-}$} & \multirow{2}{*}{$K^{-}\eta$} & \multicolumn{1}{c}{} & & \\
			 \cline{4-6}
	&	&	&   \multicolumn{2}{c||}{\multirow{3}{*}{  $K^{-}$ }} & \multirow{3}{*}{ $K^{\ast -}\,(K^{-}\pi^{0}+\overline{K}^{0}\pi^{-})$}   \\
\cline{2-3}
 & \multirow{2}{*}{ $K^{-}\pi^{0}$	} & \multirow{2}{*}{$K^{-}\eta^{\prime}$	} & \multicolumn{1}{c}{} &  &  \\
	&	&	&	\multicolumn{1}{c}{}&	&	\\
\cline{2-6}
 & \multirow{2}{*}{$\pi^{0} \pi^{+}$}  & \multirow{2}{*}{$\pi^{+}K^{0}$ } & \multicolumn{2}{c||}{\multirow{3}{*}{$\pi^{+}$ }} & \multirow{3}{*}{$\rho^{+}\,(\pi^{0} \pi^{+})$ }   \\
	&	&	&	\multicolumn{1}{c}{}&	&	\\
\cline{2-3} 
 & \multirow{2}{*}{$ K^{+} \overline{K}^{0}$ } & \multirow{2}{*}{$\eta K^{+}$ } & \multicolumn{1}{c}{}& & \\
 			\cline{4-6}
	&	&	 & \multicolumn{2}{c||}{\multirow{3}{*}{$K^{+}$}}&  \multirow{3}{*}{$K^{\ast +}\,(\pi^{0}K^{+}+\pi^{+}K^{0})$ }  \\
\cline{2-3}  
 & \multirow{2}{*}{$\pi^{0} K^{+}$} & \multirow{2}{*}{$\eta^{\prime} K^{+}$  } &  \multicolumn{1}{c}{}& &   \\
	&	&	&	\multicolumn{1}{c}{}&	&	\\
\hline
\end{tabular}
\end{center}
\vspace*{-0.5cm}
\caption{\label{tab::FinalHadronicStates} Final states considered in this work for the hadronic decays of the heavy neutrino in the regime $M_{N_{1}}<2.5$ GeV. For the vectorial resonances, as explained in the main text, we show (when needed) the dominant decay channels used to calculate their widths.}
\end{table}

For the hadronization of the neutral quark currents, we take the results directly from Ref.~\cite{Husek:2020fru}, shown there in Section 2.1.2. However, in that work the hadronization of the charged quark currents was not performed, hence we provide it here.
Since the HNL decays into quarks via SM-like interactions (see Eq.~\eqref{eq::IntLag}), we are left with the hadronization of the following charged quark bilinears:
\begin{alignat}{3}
\label{eq::finalh}
\left[ \, \overline{q}_i \, \gamma_{\mu} \, \gamma_5 \, q_j  \right. 
&\rightarrow \left.  P \, \right]
&&\simeq  - i \, 2 \, F \, \Omega_{A-C}^{(1)}(ij) \, p_{\mu}\, ,\nonumber \\
\left[ \,  \overline{q}_i \, \gamma_{\mu} \, q_j  \right. 
&\rightarrow \left.  P_1 \, P_2 \, \right]
&&\simeq \left[ \, 2 \, \Omega_{V-C}^{(2)}(ij)\, + \, 
 \sqrt{2} \, \frac{F_V \, G_V}{F^2} \, \sum_{V} \, \frac{s}{M_V^2 \, - \, s} \,  \Omega_{V-C}^{(1)}(ij) \, \Omega_{V-C}^{(3)} \, \right] (\, p_1\, -\, p_2\, )_{\mu} \nonumber \\
& && + \left[ \, \sqrt{2} \, \frac{F_V \, G_V}{F^2} \, (m_2^2 - m_1^2) \, \sum_{V} \, \frac{\Omega_{V-C}^{(1)}(ij) \, \Omega_{V-C}^{(3)}}{M_V^2 \, - \, s} \, \right] (\, p_1 \, + \, p_2 \,)_{\mu} \,. \nonumber \\
\end{alignat}
Here, the $\Omega$ coefficients are just numerical factors, which depend on the process considered, and whose values are given in Tables~\ref{tab:omega1a} and~\ref{tab:omega22vt} of appendix~\ref{app::2}. The information from QCD lies within the $F, F_{V}$ and $ G_{V}$ factors.
Note finally that, due to the larger uncertainties coming from other sources in this work, possible improvements in the hadronization resulting from the use of dispersive methods are not relevant and then, are not considered here.

\section{Analysis and results
\label{sect:results}}

Current knowledge of the neutrino sector allows to probe models that address the origin of neutrino masses and related features of this sector. This is generically achieved by translating the information provided by the current neutrino experimental data into constraints on the parameters of the model under study. 
In this work we intend to cover in the most general way the parameter space of the linear seesaw,  focusing on the regions where current and near-future experiments aim to explore.

For this purpose, we take the master parametrization described in Refs.~\cite{Cordero-Carrion:2018xre, Cordero-Carrion:2019qtu}, which allows to fit any Majorana neutrino mass model and automatically reproduce current experimental data. For the case of the linear seesaw, the Yukawa couplings of Eq.~\eqref{eq:mnu} are parametrized as:

\begin{equation}
\begin{split}
Y_{D}^{T} &= \left( \frac{M_{R}}{v_{\text{SM}}}\right)^{1/2}WT \left( \frac{\hat{m}_{\nu}}{v_{\text{SM}}}\right)^{1/2}U_{\ell  \nu}^{\dagger},   \\
Y_{\epsilon}^{T} &= \left( \frac{M_{R}}{v_{\text{SM}}}\right)^{1/2} W^{*}B \left( \frac{\hat{m}_{\nu}}{v_{\text{SM}}}\right)^{1/2}U_{\ell \nu}^{\dagger},
\label{eq:masterparam}
\end{split}
\end{equation}

\noindent where $v_{\text{SM}}$ is the Higgs vacuum expectation value, 
 $B=(T^{T})^{-1}(I-K)$, 
 \[
 \hat{m}_{\nu}=\text{diag}( m_{\nu_{1}}, \sqrt{\Delta m_{\text{sol}}^{2}+m_{\nu_{1}}^{2}},  \sqrt{\Delta m_{\text{atm}}^{2}+m_{\nu_{1}}^{2}} ),
 \]
and $U_{\ell \nu}$ is the neutrino (PMNS) mixing matrix. The master parametrization calculates the two Yukawa matrices as functions of the input parameters $m_{\nu_{1}}$, $U_{\ell \nu}$ and $M_{R}$ and three arbitrary matrices $W$, $T$ and $K$, which are unitary, upper triangular and antisymmetric, respectively. The role of these matrices can be viewed as follows: $W$ encloses all possible rotations in the Yukawa parameter space, while $T$ and $K$ contain the scaling of the different components of the Yukawa couplings. 
For our analysis we consider two special cases:
\begin{itemize}
    \item[] \emph{Scenario a:} we set $W= U_{\ell \nu}$,  
$T= f \times (v_{SM}/\hat{m}_{\nu})^{1/2}$ 
     and $K=0$, in such a way that one of the Yukawa matrices becomes diagonal.  Here $f$ is just a scale factor parametrizing the magnitude of the Yukawas. However, we conveniently redefine $f= \alpha 10^{-1}/f^{\prime}$ with $\alpha = (246)^{-1/2}$. 
All results in the next subsections are given in terms of $f^{\prime}$.
    Notice that $f'$ is such that $Y_{\epsilon}$ and $Y_{D}$ are proportional and inversely proportional to $f^{\prime}$, respectively.
    \item[] \emph{Scenario b:} we take a specially simple choice $W=I$, $T=g\times I$ and $K=0$, such that $Y_{D}=g^{2} Y_{\epsilon}$. Note that this parametrization leaves $Y_{D}Y_{\epsilon}^{T}$ constant and, in consequence, the neutrino mass unchanged for any value of $g$. For $g=1$, both Yukawa matrices become equal $Y_{D}=Y_{\epsilon}$ and the traditional seesaw scenario is recovered.
\end{itemize}
Both ansatzes are particularly interesting, because they provide a parameter space that, for some specific values of $f^{\prime}$ and $g$, is not only allowed by the current experimental bounds, but is also inside the measurable region of future experiments. A different choice in the parametrization structure either explores the same region or falls into non-testable or excluded regions.
For all the calculations, we set the best fit point (b.f.p.) values of $U_{\ell \nu}$, $\Delta m_{\text{sol}}$, $\Delta m_{\text{atm}}$ and mixing angles, from neutrino oscillation experiments \cite{deSalas:2017kay,deSalas:2020pgw}.

\subsection{Same-sign to opposite-sign dilepton ratio in the LSM}
\begin{figure}[hbt!]
  \centering
  \includegraphics[scale=0.7]{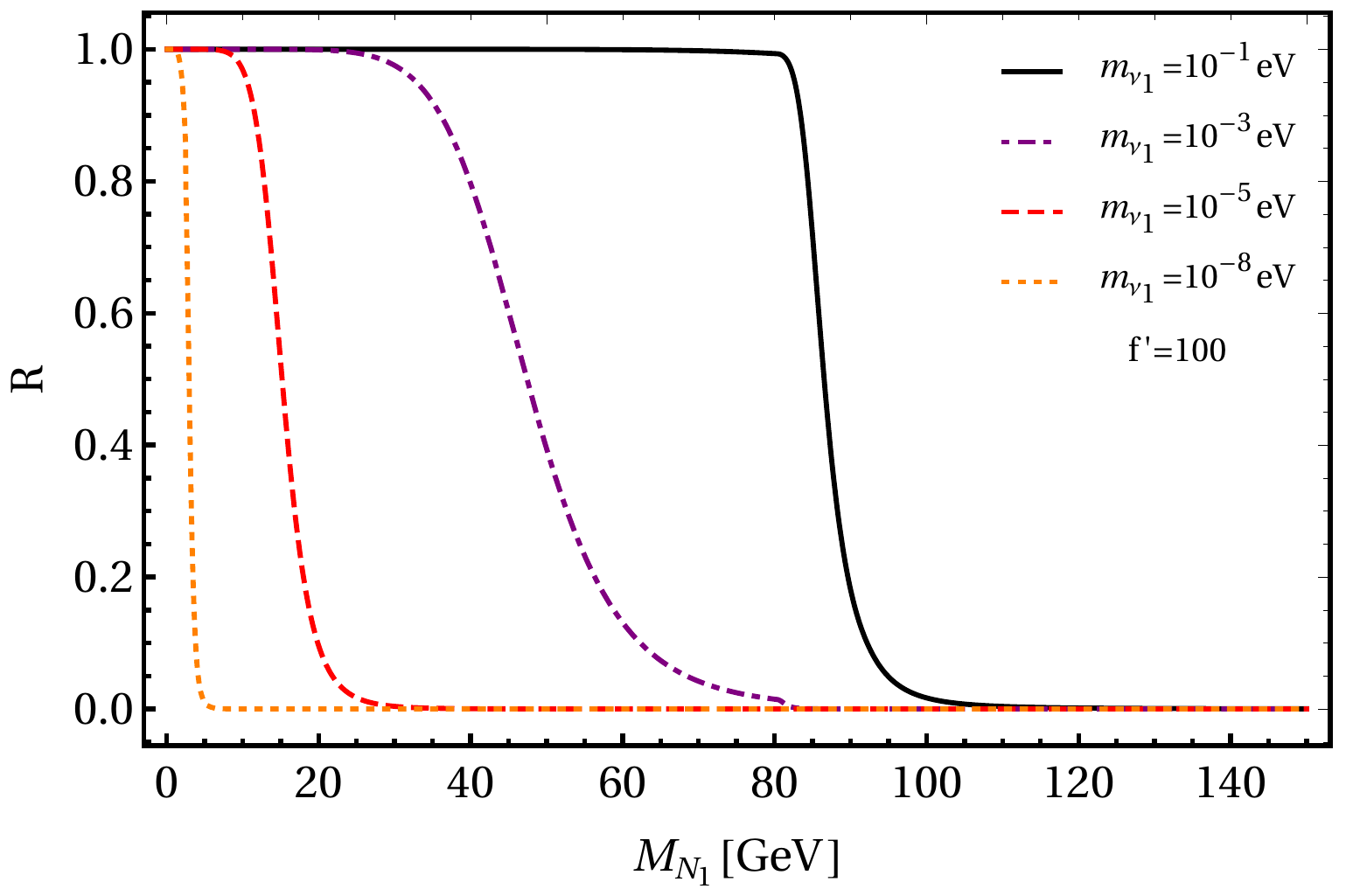}
 \caption{$R_{\ell \ell}$ vs $M_{N1}$ for different values of $m_{\nu_{1}}$.}
      \label{fig:RvsMR}
\end{figure}

From Eq.~\eqref{eq:Rll}, we can appreciate that there are two limiting cases corresponding to $R_{\ell \ell} = 1,0$. If $\Delta M \sim m_{\nu}\gg \Gamma$, the decays proceed as in the usual Majorana case and probabilities for $SS$ and $OS$ dilepton events are the same, i.e. $R_{\ell \ell}=1$. If $\Delta M \sim m_{\nu} \ll \Gamma$, the pure Dirac case is approached and then  $R_{\ell \ell}=0$. The Quasi-Dirac regime $0 < R_{\ell \ell} < 1$ occurs when $\Delta M \sim \Gamma$. 
Since $\Delta M \sim m_{\nu}$ and 
$\Gamma(M_N)$ grows quite fast with $M_N$, for smaller values of $m_{\nu_{1}}$, smaller values of $M_{N}$ are needed. This is illustrated in Fig. \ref{fig:RvsMR}, which shows $R_{\ell \ell}$ versus $M_{N_{1}}$ for different values of $m_{\nu_{1}}$. For this calculation, we have chosen \emph{Scenario a}, with a fixed value of $f^{\prime}=100$. As expected, for each specific  $m_{\nu_{1}}$, there is a relatively narrow window of $M_{N_{1}}$ values such that $0<R_{\ell \ell}<1$. For example, if $m_{\nu_{1}}=10^{-5}$ eV, then values of $10$ GeV $\lesssim M_{N_{1}} \lesssim 20$ GeV are needed in order to obtain a $R_{\ell \ell}$ value within the QD regime. Unlike the inverse seesaw model \cite{Anamiati:2016uxp}, where values of $R_{\ell \ell}<1$ are still obtained for larger values of $M_{N_{1}}$, here the current upper bound for the light neutrino mass $m_{\nu_{1}} \lesssim 0.1$ eV and a $f^{\prime}=100$ sets $R_{\ell \ell}=0$ for values of $M_{N_{1}}\gtrsim 100$ GeV.
This is why our analysis focuses on the regime of small $M_{N_{1}}$ values.

\begin{figure}[hbt!]
  \centering
  \includegraphics[scale=0.4]{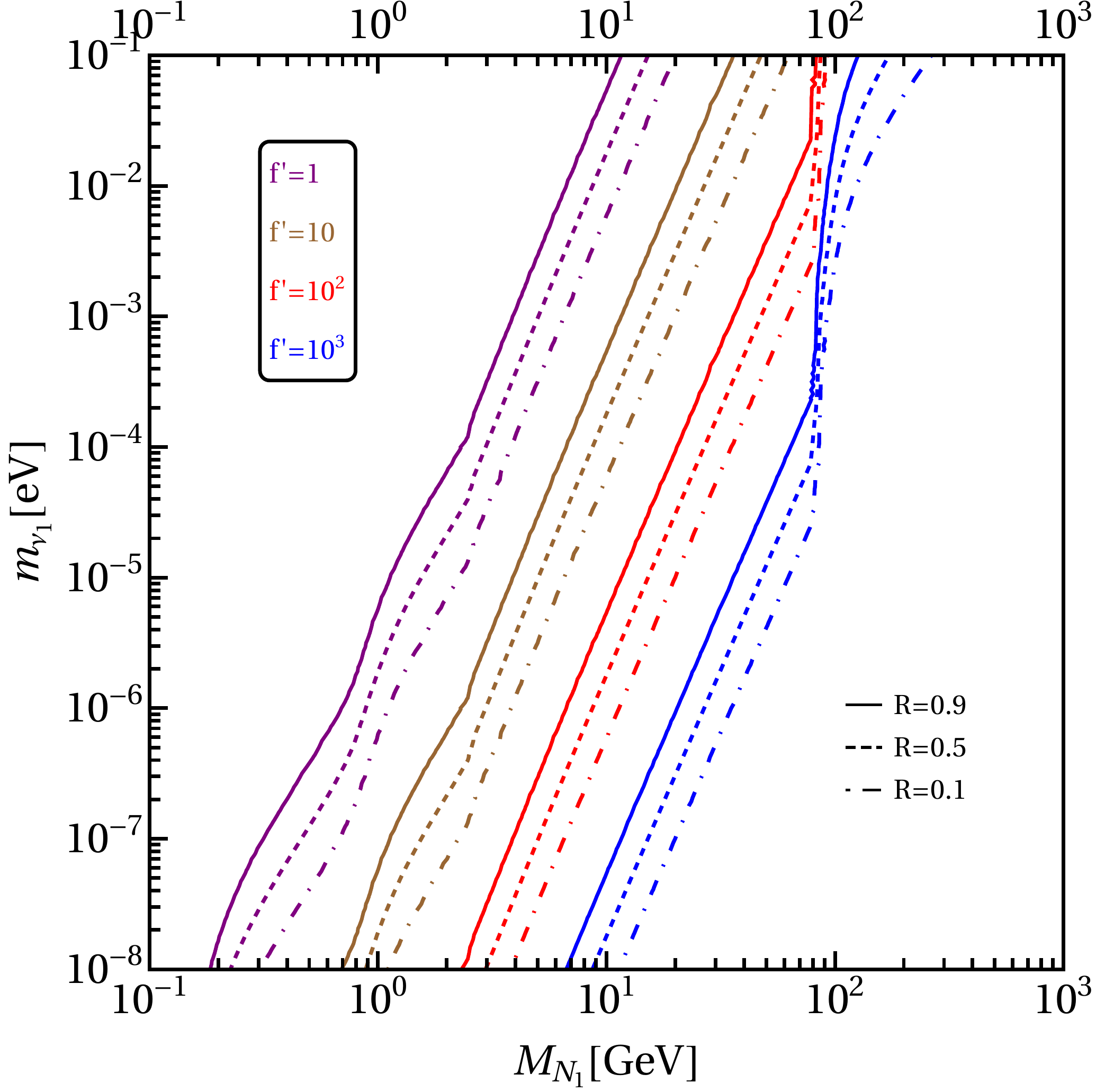}
 \caption{$m_{\nu_{1}}-M_{N_{1}}$ lines corresponding to a specific value of $R_{\ell \ell}$ and $f^{\prime}$. For each color, the three lines correspond to values of $R_{\ell \ell}=0.9$, $0.5$ and $0.1$ from left to right.}
 \label{fig:mnuvsMR}
      
\end{figure}

Another interesting plot is Fig.~\ref{fig:mnuvsMR}, which shows the regions in the $m_{\nu_{1}}$-$M_{N_{1}}$ plane that belong to the QD regime (i.e. $0< R_{\ell \ell} <1$) in the linear seesaw model. 
This is done within \emph{Scenario a} for some specific values of $f^{\prime}$. The different colors magenta, brown, red and blue correspond to values of $f^{\prime}=1$, $10$, $10^{2}$, and $10^{3}$, respectively (recall that $Y_\epsilon$ grows with $f'$). In this Figure, for each $f^{\prime}$ there are three lines corresponding to values of $R_{\ell \ell}= 0.9$, $0.5$ and $0.1$, from left to right. Therefore, for a specific  $m_{\nu_{1}}$, each colored band roughly provides a range of $M_{N_{1}}$ values for which  $0 < R_{\ell \ell} < 1$. Note though that this QD regime is a continuum: the upper-left corner represents the Majorana case, while the lower-right corner approaches the Dirac limit. This QD-$M_{N_{1}}$ window of values moves to the left as $m_{\nu_{1}}$ becomes smaller. For example, for $f^{\prime}=10$ and $m_{\nu_{1}}=10^{-3}$ eV, the heavy neutrinos become Quasi-Dirac for values of $10$ GeV $\lesssim M_{N_{1}} \lesssim$ $20$ GeV, while if $m_{\nu_{1}}=10^{-6}$ eV, then the QD behaviour occurs for $2$ GeV $\lesssim M_{N_{1}} \lesssim$ $4$ GeV. These specific ranges can be additionally constrained from the current and expected experimental bounds on the mixing $U_{N_{1}e}$. Fig~\ref{fig:mnuvsMR} allows us to directly translate those constraints into the QD regime, as we study in the next section.

\subsection{Heavy to light neutrino mixing $U_{N \ell} $}

\noindent Here, we analyze the $|U_{N_{1}\ell}|^{2} - M_{N_{1}}$ region and identify the zones in the parameter space where some current experiments could have sensitivity; the prospects of future experiments are also considered. The numerical analysis is based on the systematic diagonalization of the $9 \times 9$ mass matrix of the neutral states $M_{\nu}$ (see Eq.~\eqref{eq:numatrix}) 

\begin{equation}
M_{\nu}= U \hat{M}_{\nu} U^{T}\, ,
\end{equation}
with $U$ containing 
the PMNS mixing matrix and the heavy-light neutrino mixing elements $U_{N_{1} \ell}$ as well.  For the $U$ calculation, we parametrize the Yukawa couplings as in Eq.~\eqref {eq:masterparam}. Therefore, after fixing the neutrino oscillation parameters to the best fit point, we obtain numerical expressions of the mixings $U_{N_{1} \ell}$ as a function of $M_{N_{1}}$ and $f^{\prime}$ or $g$, depending on the specific scenario. As we sketched in Section~\ref{sect:DecayRates}, for the sake of simplicity we considered a large decoupling on the masses of the second and third generation of the heavy neutrinos with respect to $M_{N_{1}^{(\prime)}}$, i.e. $M_{N_{1}^{(\prime)}} \ll M_{N_{2}^{(\prime)}} , 
M_{N_{3}^{(\prime)}}$.\footnote{ Due $M_{N_{1}} \sim M_{N_{1}^{\prime}}$, the results and figures involving $M_{N_{1}}$ apply also for $M_{N_{1}^{\prime}}$.} However, for completeness, at the end of this section we briefly comment on the phenomenologically different inverted-hierarchical case, where the mass of the third-generation heavy neutrino pair is lighter than the other two pairs.

\begin{figure}
  \centering
  \includegraphics[scale=0.9]{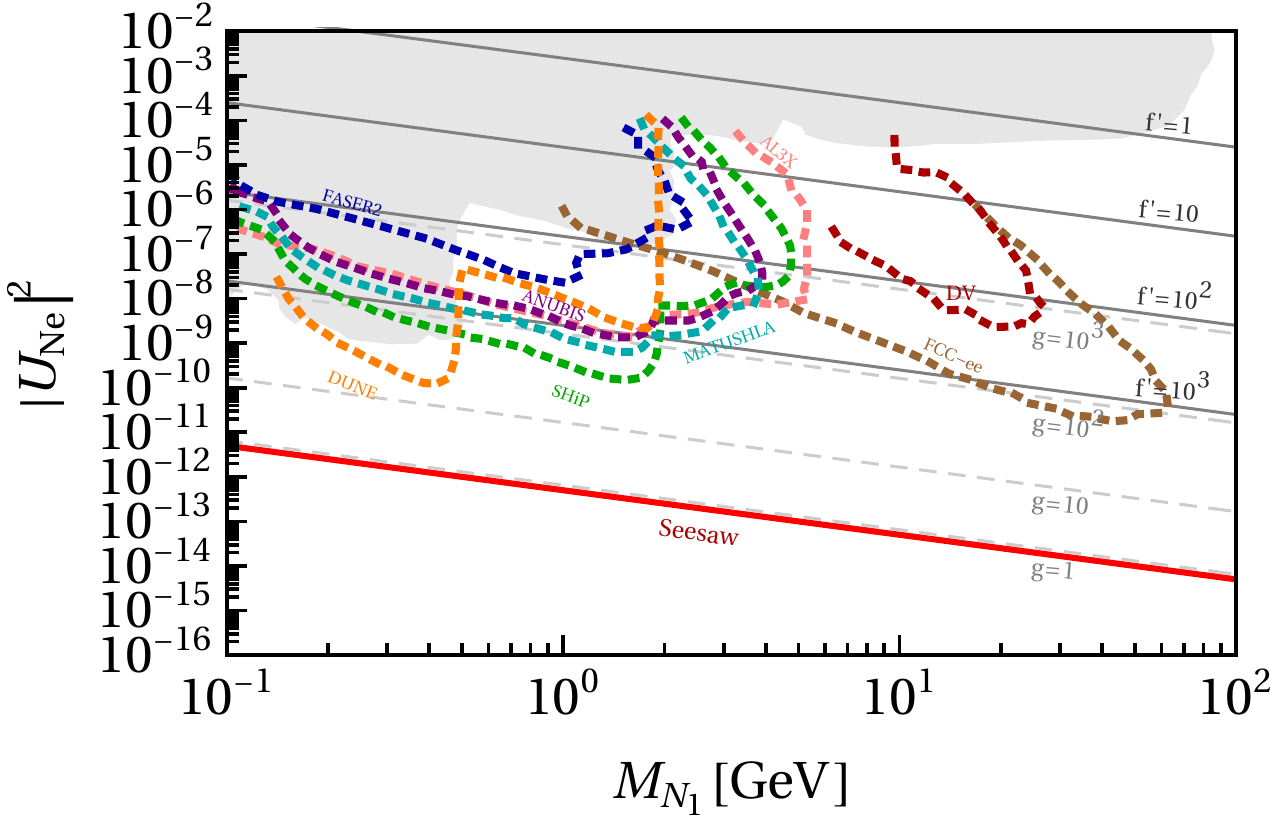}
 \caption{Active-sterile neutrino mixing $|U_{N_{1}e}|^{2}$ versus the neutrino mass $M_{N_{1}}$, for different values of the parameters $f^{\prime}$ and $g$. Dashed lines correspond to the projected sensitivities of future experiments, while the gray shaded area indicates the regions already excluded from current searches as depicted in Ref.~\cite{Bolton:2019pcu}. The dashed dark red line limits the zone of the sensitivity of DV searches. The solid red line represents the vanilla seesaw model.}
      \label{fig:UvsMR}
\end{figure}

Figure \ref{fig:UvsMR} shows $|U_{N_{1}e}|^{2}$ versus the heavy neutrino mass $M_{N_{1}}$, for different values of $f^{\prime}$ and $g$: solid and dashed gray straight lines correspond to scenarios \emph{a} and \emph{b} respectively. In \emph{Scenario a} all light neutrino masses $m_{\nu_{i}}$ enter in all $Y_{\epsilon} $ entries, while $Y_{D}$ is independent of the light neutrino masses. Obviously, because $m_{\nu_{3}} \gg m_{\nu_{2}} \gg m_{\nu_{1}}$, the mixings $U_{N_{1}\ell}$ will not depend on $m_{\nu_{1}}$. Otherwise, in \emph{Scenario b}, each $M_{R_{i}}$ is responsible for each $m_{\nu_{i}}$, therefore an explicit dependence on $m_{\nu_{1}}$ is expected. For our calculations on \emph{Scenario b}, we set $m_{\nu_{1}}=10^{-3}$ eV. The current constraints and some future projections on the mixing are shown by the shaded region and the colored dashed lines. We can appreciate that for small (large) values of $f^{\prime}$ ($g$), the mixing $U_{N_{1}e}$ becomes strongly constrained by the experimental bounds in a wide range of $M_{N_{1}}$ values. This means that, if there is some appreciable hierarchy between the Yukawa couplings $Y_{D}$ and $Y_{\epsilon}$ (see the description of both scenarios above), the predicted mixing falls into the range testable by present and near-future experiments. For some values of $g\lesssim 20$ the model remains unconstrained by the low energy experimental bounds. A value of $g=1$, where $Y_{D} = Y_{\epsilon}$, corresponds to the traditional  seesaw mechanism \cite{PhysRevLett.44.912,GellMann:1980vs,PhysRevD.22.2227} where the heavy-light neutrino mixing is given by $U_{N_{1}e} \simeq \sqrt{m_{\nu_{1}}/M_{N_{1}}}$. It is interesting to note that the traditional seesaw represented by the red  line (for which we set $m_{\nu_{1}}=10^{-3}$) is not sensitive to any experimental bound, while our model could be strongly constrained in a large part of the parameter space. The projected sensitivities for ANUBIS (purple dashed line), MATHUSLA (cyan dashed line), SHiP (green dashed line), DUNE (orange dashed line), FASER2 (blue dashed line), FCC-ee (brown dashed line) and AL3X (pink dashed line) are taken from Refs.~\cite{deVries:2020qns,Hirsch:2020klk,Helo:2018qej,Bolton:2019pcu}, while the already excluded values (grey shaded region) are taken from Ref.~\cite{Deppisch:2015qwa,Bolton:2019pcu}. There is in addition a region in the $|U_{N_{1}e}|^{2}-M_{N_{1}}$ plane where a displaced vertex search could have sensitivity: the region limited by the darker red dashed line represents a 95 \% CL reach at $\sqrt{s}=13$ TeV of a multitrack displaced vertex strategy described in Ref.~\cite{Cottin:2018nms}.  As depicted in this figure, for values of $f^{\prime}$ around $10 \lesssim f^{\prime} \lesssim 10^{2}$ and a specific window of $M_{N_{1}}$ values, a discovery of a displaced vertex signal in the model would be possible at the high-luminosity ($\mathcal{L}=3000 fb^{-1}$) LHC. A detailed study of displaced vertex signals for QD neutrinos in the linear seesaw model goes beyond the scope of this work, but it could be worth pursuing. Note finally that both scenarios \emph{a} and \emph{b} present a symmetry axis of $U_{N_{1}e}$, in powers of the parametrization factor, for $f^{\prime} \sim 10^{4.5}$ and $g = 10^{0} $, respectively.

The characterization of the QD regime through the $R_{\ell \ell}$ observable implies a possibility for the experiments to measure the charge of the two final leptons (see Section~\ref{sect:sum}). Nevertheless, some of the aforementioned low energy scale experiments, except the ones concerning to LHC-displaced vertex searches, will not be able to differentiate the charge and then will be blind to this observable. However, the bounds on the mixing themselves can place already stringent constraints into the Quasi-Dirac regime, which can be found by studying the interplay between Fig.~\ref{fig:mnuvsMR} and~\ref{fig:UvsMR}.  According to Fig.~\ref{fig:mnuvsMR}, in \emph{Scenario a} each value of  $m_{\nu_{1}}$ leads to a specific $M_{N_{1}}$ range where  $0<R_{\ell \ell}<1$. At the same time, each $M_{N_{1}}$-QD range gets extra constraints by the experimental bounds shown in Fig.~\ref{fig:UvsMR}. For instance, with $f^{\prime}=10$ and $m_{\nu_{1}}=10^{-3}$ eV, the corresponding QD masses $10$ GeV $\lesssim M_{N_{1}} \lesssim$ $20$ GeV (see Fig.~\ref{fig:mnuvsMR}) fall into the region where displaced vertex searches could have sensitivity, whereas for a value of $m_{\nu_{1}}=10^{-6}$ eV, some region of the QD masses could be excluded by some of the low mass experiments. On the other hand, for $f^{\prime}=10^{3}$, the FCC-ee will be probing masses between $20$ GeV $\lesssim M_{N_{1}} \lesssim$ $60$ GeV, so it will be sensitive to QD neutrinos for $10^{-7}$ eV $\lesssim m_{\nu_{1}} \lesssim$ $10^{-4}$ eV, but it is not until one reaches very low values of $m_{\nu_{1}}$ that any foreseen low-energy experiment could be sensitive to QD neutrinos. These observations allow us to conclude that, for a specific value of $f^{\prime}$, the QD regime gets stronger constraints as the light neutrino mass becomes smaller. For \emph{Scenario b}, since in this case the mixing depends on both the heavy and the light neutrino masses (and then also on the decay width), instead of the fixed-value $R_{\ell \ell}$ lines depicted in Fig.~\ref{fig:mnuvsMR}, we find straight vertical lines [from Eq.~\eqref{eq:Rll}]. This means that for a given $g$ and $M_{N_{1}}$, all values of $m_{\nu_{1}}$ give an approximately equal $R_{\ell \ell}$, i.e. $R_{\ell \ell}$ is independent of the lightest neutrino mass. Therefore, there is more freedom when translating the bounds on the mixing to the Quasi-Dirac regimes, and in consequence the model becomes less constrained.

The inverted hierarchical case, with the third-generation heavy neutrino being the lightest of its kind, i.e. $M_{N_{3}^{(\prime)}} \ll M_{N_{1}^{(\prime)}} , M_{N_{2}^{(\prime)}}$, exhibits some different features, which we comment here briefly. First, it mixes predominantly into the $\tau$ neutrino, then we should compare $U_{N_{3} \tau}$ against the previous $U_{N_{1} e}$. For \emph{Scenario a}, no relevant difference is observed, but \emph{Scenario b} is more interesting: now $U_{N_{3} \tau}$ is $m_{\nu_{1}}$-independent, but $m_{\nu_{3}}$-dependent. Similar to the $g=1$ traditional seesaw scenario, where $U_{N_{3} \tau} \sim \sqrt{m_{\nu_{3}}/M_{N_{3}}}$, all dashed gray curves in Fig.~\ref{fig:UvsMR} move upwards with respect to the previous case. All this leads for $R_{\ell \ell}$ to a similar QD behaviour as presented in Fig.~\ref{fig:mnuvsMR}, thus allowing for stronger constraints on the QD regime.

Let us now briefly mention the effects of current lepton-flavour violating (LFV) bounds in our model. We want to add that we have calculated the branching ratio of the $\mu \rightarrow e \gamma $ process in both scenarios. Taking into account that our model does not have extra charged scalars, the only contribution to this LFV process comes from the loop mediated by the $W$ boson.  We have found that $Br(\mu \rightarrow e \gamma)$ can provide additional relevant constraints only for extreme values of the $f^{\prime}$ and $g$, which lie on the already excluded area by the experimental searches depicted in Fig.~\ref{fig:UvsMR}.

Finally, regarding neutrinoless double beta decay ($0\nu \beta \beta$) searches, following Ref.~\cite{Bolton:2019pcu} one can see that the corresponding bounds get worse as the mass splitting between the pair of heavy neutrinos becomes smaller --- as it is expected since $0\nu \beta \beta$ is a LNV process that vanishes away for $\Delta M = 0$, i.e. in the Dirac case. In the linear seesaw model ($\Delta M_{i} = m_{\nu_{i}}$) we deal with values of $\Delta r \equiv \Delta M_{1} /M_{N_{1}}\in [10^{-11},10^{-19}]$, which are farther above the worst upper-bound $\Delta r =10^{-6}$ depicted in Figs.~9 and 10 of Ref.~\cite{Bolton:2019pcu}. Therefore, we conclude that the constraints coming from $0\nu \beta \beta$ are not competitive in the QD region of interest for our work.

\section{Conclusions\label{sect:conc}}

We studied the Quasi-Dirac nature of the heavy neutrinos in the most minimal version of the linear seesaw. This model contains two extra heavy neutrinos, each replicated in three generations. The mass splitting of the heavy neutrino pair in each generation is shown to be of the order of the light-neutrino masses, as shown in Eq.~\eqref{eq:splitt}. An interesting way to study the nature of these additional neutrinos is through the ratio $R_{\ell \ell}$, which relates the number of same-sign to opposite-sign dilepton final states 
(in $\ell\ell j j$ events at hadron colliders)
 when a heavy neutrino is involved, and may take any value from 0 (Dirac) to 1 (Majorana), with the intermediate region displaying the QD behaviour. We showed that in the LSM this observable is controlled by both the masses of the light neutrinos and the decay widths of their heavy partners (see Eq.~\eqref{eq:Rll}). Therefore, focusing on the first generation and decoupling it from the other two, we computed its decay width, paying special attention to the quark modes within the non-perturbative regime of $\text{E}_{\text{QCD}} \lesssim 2.5$ GeV. In order to properly hadronize these quark currents, we employed chiral perturbation theory and its higher energy extension the resonance chiral theory, which allowed us to include all features of QCD and systematically calculate all the variety of final hadronic channels.

We performed a numerical analysis based on the systematic diagonalization of the full mass matrix of the neutral states (see Eq.~\eqref{eq:numatrix}). Due to the richness of the model, we implemented the master parametrization, that allowed us to scan the parameter space of the model, while automatically fit the current data of the neutrino sector in the most general way. Within this general framework, we choose two simplified realizations of the Yukawas (scenarios \emph{a} and \emph{b}) 
that, nonetheless, embraced the most interesting region of the model.

Unlike other seesaw models with QD regimes, in the linear seesaw the pair of heavy neutrinos exhibits a Quasi-Dirac behavior for relatively low masses.

For each scenario we found the regions of heavy neutrino masses  which exhibit the QD regime, i.e. $0 < R_{\ell \ell}<1$, and which at the same time are consistent with the current bounds for the light neutrino mass: 

 \emph{Scenario a} showed a dependence on $m_{\nu_{1}}$, such that lower light-neutrino masses entailed lower masses for the QD heavy neutrinos as well, in the range of a few GeV's; these QD regions in \emph{Scenario b} on the other hand, happen to be $m_{\nu_{1}}$-independent.  

A phenomenological analysis using experimental data on current and near-future experiments was also carried out to constraint ---within both scenarios--- the heavy-light neutrino mixing $U_{N_{1}e}$ as a function of the heavy-neutrino mass $M_{N_{1}}$ (see Fig.~\ref{fig:UvsMR}). We found that low-energy experiments such as SHiP, MATUSHLA, ANUBIS, DUNE, FASER2 and AL3X together with displaced vertex searches and prospects for the FCC-ee, will explore a large part of the parameter space for both scenarios, either discovering or placing stringent bounds on the parameter space of the LSM, which is still far from reach for the ordinary seesaw. Therefore, we concluded that current and near-future experiments are actually probing hierarchical Yukawas, with the equal-Yukawa case remaining unbounded.

Despite the stated difficulty in measuring directly $R_{\ell \ell}$, we could translate the previous mixing constraints into bounds on the Quasi-Dirac nature of the heavy neutrinos through this observable, what was achieved by comparing Figs.~\ref{fig:mnuvsMR} and~\ref{fig:UvsMR}. By doing so, we found that the QD regime is more strongly constrained in  \emph{Scenario a} than in  \emph{Scenario b}. In the former, the lower the light-neutrino mass, the lower the mass range yielding $R \neq 0,1$ and then, the more constrained the QD regime turns out to be. Accordingly, this would lead to a Dirac behavior for these heavy neutrinos. Instead,  \emph{Scenario b} was shown to be less constrained: the QD mass regime was shifted to larger masses and the $m_{\nu_{1}}$ dependence of the mixing allowed to take it away of the exclusion area. The inverted hierarchical case was also considered and some of its main features were briefly discussed.

Finally, the most up-to-date stringent bounds on the neutrinoless double beta decay process as well as the lepton-flavor violating process $\mu \rightarrow e \gamma$ were also addressed. We concluded that the current experimental upper limits on these processes do not yield competitive constraints as compared to the ones shown in Fig.~\ref{fig:UvsMR}.

\hfill

\centerline{\bf Acknowledgements}

\medskip

We are grateful to Martin Hirsch for the very helpful discussions. We wish to thank Jorge Portolés for the fruitful discussions on the hadronization of the quark currents. This work was supported in part by FONDECYT (Chile) grants No. 11180722, 1180232 and 1170171, and by ANID (Chile) PIA/APOYO AFB 180002, as well as by Grants No. FPA2017-84445-P and SEV-2014-0398 (AEI/ERDF, EU) and by PROMETEO/2017/053 (GV).

\appendix

\renewcommand{\theequation}{\Alph{section}.\arabic{equation}}
\renewcommand{\thetable}{\Alph{section}.\arabic{table}}
\renewcommand{\thefigure}{\Alph{section}.\arabic{figure}}

\let\appsect\section
\renewcommand{\section}{
\setcounter{equation}{0}
\setcounter{table}{0}
\setcounter{figure}{0}
\appsect}

\section*{Appendices}

\section{\texorpdfstring{\boldmath $\Omega$}{Omega} coefficients in Eq.~\eqref{eq::finalh}}
\label{app::2}
In Tables~\ref{tab:omega1a} and~\ref{tab:omega22vt} we present the values of the $\Omega$ coefficients found in Eq.~\eqref{eq::finalh}, for the different final and intermediate-state contributions. Note that the quark-type symbols used in the table header as $ud$, stand for the field content of the quark bilinear under hadronization, while the symbols representing mesons --- either pseudoscalar or vectorial resonances --- stand for the actual physical states. We define $\sin \theta_{P}\equiv s_{P}$ and $\cos \theta_{P}\equiv c_{P}$, with $\theta_{P}$ the mixing angle between the octet $(\eta_{8})$ and singlet $(\eta_{0})$ strong-interaction eigenstates of the pseudoscalar meson multiplet giving the $\eta$ and $\eta^{\prime}$ physical states, taken to be $\theta_{P}=- 20^{\circ}$~\cite{HerreraSiklody:1997kd,Kaiser:1998ds}.
\begin{table}[h!]
\begin{center}
\renewcommand{\arraystretch}{1.5}
\begin{tabular}{|c|c|c|c|c|}
\cline{2-5}
\multicolumn{1}{c|}{}& \multicolumn{4}{c|}{$\Omega_{A-C}^{(1)}(ij)$,\; $-\Omega_{V-C}^{(1)}(ij)$}  \\
\hline
$P$ \hfill$\Big|$\hfill $V$ & $ud$ & $du$ & $us$ & $su$ \\
\hline
\hline
$\pi^+$ \hfill$\Big|$\hfill $\rho^{+}$ & $\frac{1}{\sqrt{2}}$ & 0 & 0 & 0   \\
\hline
$\pi^-$ \hfill$\Big|$\hfill $\rho^{-}$ & 0  & $\frac{1}{\sqrt{2}}$ & 0 & 0 \\
\hline
$K^+$ \hfill$\; \,  \Big|$\hfill $ K^{\ast +}$ & 0 & 0 & $\frac{1}{\sqrt{2}}$ &  0 \\
\hline
$K^-$ \hfill$\; \, \Big|$\hfill $ K^{\ast -}$ & 0  & 0  & 0 & $\frac{1}{\sqrt{2}}$  \\
\hline
\end{tabular}
\end{center}
\vspace*{-0.5cm}
\caption{\label{tab:omega1a} Factors $\Omega_{A-C}^{(1)}(ij)$ and $\Omega_{V-C}^{(1)}(ij)\, $.
}
\end{table}
\begin{table}[h!]
\begin{center}
\renewcommand{\arraystretch}{1.5}
\begin{tabular}{|c||c|c|c|c|}
\cline{2-5}
\multicolumn{1}{c|}{} & \multicolumn{4}{c|}{$-2\Omega_{V-C}^{(2)}(ij)$,\;\;  $\Omega_{V-C}^{(3)}$}  \\
\hline
$P_1 P_2$ & $ud$ \hfill$\Big|$\hfill $\rho^{+}$  & $du$ \hfill$\Big|$\hfill $\rho^{-}$ & $us$ \hfill$\Big|$\hfill $ K^{\ast +}$ & $su$  \hfill$ \Big|$\hfill $ K^{\ast -}$\\
\hline
\hline
$\pi^0 \pi^+$ & $\sqrt{2}$ & 0 & 0 & 0 \\
\hline
$\pi^- \pi^0$  & 0 & $\sqrt{2}$  & 0 & 0  \\
\hline
$K^+ \overline{K}^0$ & 1 & 0 & 0 & 0 \\
\hline
$K^0 K^-$   & 0 & 1 & 0 & 0  \\
\hline
$\pi^0 K^+$  & 0 & 0 & $\frac{1}{\sqrt{2}}$ & 0 \\
\hline
$K^- \pi^0$ & 0 & 0 & 0 & $\frac{1}{\sqrt{2}}$  \\
\hline
$ \pi^+ K^0$ & 0 & 0 & 1 & 0 \\
\hline
$\overline{K}^0 \pi^-$ & 0 & 0 & 0 & 1\\
\hline
$\eta  K^+  $  & 0 & 0 & $\sqrt{\frac{3}{2}}c_{P}$ & 0\\
\hline
$K^- \eta $  & 0 & 0 & 0 & $\sqrt{\frac{3}{2}}c_{P}$ \\
\hline
$ \eta' K^+ $  & 0 & 0 & $\sqrt{\frac{3}{2}}s_{P}$ & 0  \\
\hline
$K^- \eta' $  & 0 & 0 & 0 & $\sqrt{\frac{3}{2}}s_{P}$  \\
\hline
\end{tabular}
\end{center}
\vspace*{-0.5cm}
\caption{\label{tab:omega22vt} Factors $\Omega_{V-C}^{(2)}(ij)$ and $\Omega_{V-C}^{(3)}\,$. 
}
\end{table}
\clearpage

\providecommand{\href}[2]{#2}\begingroup\raggedright\endgroup

\end{document}